\documentclass[useAMS,usenatbib]{mn2e}
\usepackage{graphicx}
\usepackage{color}
\usepackage{ulem}
\def\apj{ApJ}
\def\apjs{ApJS}

\def\aap{A\&A}
\def\aaps{A\&AS}
\def\aj{AJ}
\def\mnras{MNRAS}

\def\pasp{PASP}

\def\araa{Ann.Rev.Astron.Astrophys.}

\title[An extreme dwarf galaxy J1046$+$4047]
{J1046$+$4047: an extremely low-metallicity dwarf star-forming galaxy with O$_{32}$ = 57}
\author[Y. I. Izotov et al.]{Y. I.\ Izotov$^{1}$\thanks{Corresponding author: yizotov@bitp.kiev.ua},
T. X.\ Thuan$^{2,3}$ and N. G.\ Guseva$^{1}$\\
                $^{1}$Bogolyubov Institute for Theoretical Physics,
                     National Academy of Sciences of Ukraine,
                     14-b Metrolohichna str., Kyiv, 03143, Ukraine,\\
                $^{2}$Astronomy Department, University of Virginia, 
                     P.O. Box 400325, Charlottesville, VA 22904-4325,\\
                 $^{3}$Institut d'Astrophysique de Paris (UMR 7095 CNRS \& Sorbonne Universit\'e),98 bis Bd Arago, F-75014, Paris, France\\
}
\begin{document}


\pagerange{\pageref{firstpage}--\pageref{lastpage}} \pubyear{2023}

\maketitle

\label{firstpage}

\begin{abstract}
Using the optical spectrum obtained with the Kitt Peak Ohio State Multi-Object
Spectrograph (KOSMOS) mounted on the Apache Point Observatory (APO) 3.5m
Telescope and the Sloan Digital Sky Survey (SDSS) spectrum, we study the properties of one of the most 
metal-poor dwarf star-forming galaxies (SFG) in the local Universe, 
J1046$+$4047. The galaxy, with a redshift $z$=0.04874, was selected from the
Data Release 16 (DR16) of the SDSS. 
Its properties are among the most extreme for SFGs in several ways. The oxygen
abundance 12~+~log(O/H) = 7.082$\pm$0.016 in J1046$+$4047 is among the lowest
ever observed. With 
an absolute magnitude $M_g$ = $-$16.51 mag,
a low stellar mass $M_\star$ = 1.8$\times$10$^6$~M$_\odot$ and a very low
mass-to-light ratio $M_\star$/$L_g$~$\sim$~0.0029 (in solar units),
J1046$+$4047 has a very high specific star-formation rate
sSFR $\sim$ 430 Gyr$^{-1}$, indicating very active ongoing star formation. 
Another striking feature of J1046$+$4047 is that it possesses a ratio O$_{32}$ = 
$I$([O~{\sc iii}]$\lambda$5007)/$I$([O~{\sc ii}]$\lambda$3727) $\sim$ 57. 
Using this extremely high O$_{32}$, we have confirmed and improved the 
strong-line calibration for the 
determination of oxygen abundances in the most metal-deficient galaxies, 
in the range 12~+~log(O/H)~$\la$~7.65. This improved method is applicable
for all galaxies with O$_{32}$ $\leq$ 60.
We find the H$\alpha$ emission line in J1046$+$4047 to be enhanced by
some non-recombination processes and thus can not be used for the determination
of interstellar extinction.  
\end{abstract}


\section{Introduction}\label{sec:INT}

The importance of finding of extremely metal-deficient (XMD) low-redshift
star-forming galaxies (SFGs), with oxygen abundances
12~+~log(O/H) in the range $\sim$ 6.9 -- 7.1, is emphasized by
the fact that they share many of the same properties with the dwarf galaxies at 
high redshifts (low metallicities and luminosities, high star formation rates
and high specific star formation rates) and thus may be considered as their
best local counterparts. These high-redshift dwarfs are thought to be
responsible for the reionization of the Universe and so it is crucial to study
them in detail. 
However, until very recently, no galaxy with
a definite oxygen abundance below 12~+~log(O/H) $\sim$ 7.1 is known at high redshifts, even if recent {\sl James Webb Space Telescope} ({\sl JWST}) observations of galaxies with $z$ larger than $6$ are included \citep*{O09,WC09,Y11,M13,B15,I18c}. Only lately, \citet{A23} have reported the discovery by {\sl JWST} of eight
SFGs at $z$ $\sim$ 6--8 with 12 + log(O/H) approaching the value of $\la$ 7.0
and stellar masses of $\la$~10$^7$M$_\odot$. These properties are similar to
those of the most metal-deficient low-$z$ SFGs considered in this paper.

Because of the lack of many known XMDs in the early Universe, our strategy is to
search for and use low-$z$ metal-poor SFGs as their local proxies. 
A few such XMD objects have been found 
at low redshifts. \citet{H16} have 
reported 12~+~log(O/H) = 7.02$\pm$0.03 for the galaxy AGC~198691. 
\citet{I18a}, \citet{Iz19a} and \citet{Iz21} have derived oxygen abundances in
J0811$+$4730, J1234$+$3901 and J2229$+$2725
of 6.98$\pm$0.02, 7.035$\pm$0.026 and 7.085$\pm$031, respectively. 
All these galaxies were selected from the Sloan Digital Sky Survey 
(SDSS). \citet{Ko20} have recently reported the discovery of the galaxy 
J1631$+$4426 with 12~+~log(O/H) = 6.90$\pm$0.03. However, \citet*{T22} have derived a
higher abundance 12~+~log(O/H) = 7.14$\pm$0.03 for this galaxy, based on their own  
independent observations. In all these studies, the 
oxygen abundances have been derived by the direct $T_{\rm e}$ method from
spectra of the entire galaxy. Common
characteristics of these XMD galaxies are their low stellar mass
$M_\star$~$\la$~10$^7$~M$_\odot$ and a very compact morphological structure. 
In addition to these galaxies, very low oxygen abundances of 7.01$\pm$0.07, 
6.98$\pm$0.06 and 
6.86$\pm$0.14 have also been found by \citet{I09} in three individual H~{\sc ii}
regions of the XMD SBS~0335$-$052W. A value of 
6.96$\pm$0.09 has been obtained by \citet{An19} in one of the H~{\sc ii} 
regions in the dwarf irregular galaxy DDO~68. However, other regions in 
SBS~0335$-$052W and DDO~68 do show oxygen abundances above 7.1.

In this paper, we present new spectroscopic observations of the XMD galaxy J1046$+$4047, obtained with the Kitt Peak Ohio State Multi-Object Spectrograph (KOSMOS), mounted on the Apache Point Observatory (APO) 3.5m Telescope. We supplement the KOSMOS observations with SDSS spectroscopic observations. 
The galaxy was selected from the SDSS Data Release 16 (DR16) data base 
\citep{Ah20} because its line ratios suggest it to be an extremely low-metallicity object.
It also drew our attention because of its extremely 
high (extinction-corrected) O$_{32}$ value of 57, defined as the ratio
$I$([O~{\sc iii}]$\lambda$5007)/$I$([O~{\sc ii}]$\lambda$3727). 
This high ratio 
indicates that the galaxy may contain density-bounded H~{\sc ii} regions, and 
thus be a potential Lyman continuum leaker \citep[e.g. ][]{JO13} 
and a good local counterpart to the high-redshift dwarf galaxies.
Its coordinates, redshift and other characteristics
obtained from the photometric and spectroscopic SDSS and the {\sl Galaxy Evolution 
  Explorer} ({\sl GALEX}) data bases are presented in Table \ref{tab1}. We note
that the galaxy was not detected by space {\sl Wide-field Infrared Survey
Explorer} ({\sl WISE}) because in part of its low metallicity and thus low dust
content.

\begin{table}
\caption{Observed and derived characteristics of J1046$+$4047 \label{tab1}}
\begin{tabular}{lr} \hline
Parameter                 &  J1046$+$4047       \\ \hline
R.A.(J2000)               &  10:46:09.23 \\
Dec.(J2000)               & +40:47:07.07 \\
  $z$                     &  0.04874$\pm$0.00006     \\
{\sl GALEX}  FUV, mag     &   20.27$\pm$0.24      \\
{\sl GALEX}  NUV, mag     &   21.19$\pm$0.28      \\
SDSS $u$, mag             &   20.85$\pm$0.06      \\
SDSS $g$, mag             &   20.18$\pm$0.02      \\
SDSS $r$, mag             &   21.39$\pm$0.07      \\
SDSS $i$, mag             &   20.71$\pm$0.04      \\
SDSS $z$, mag             &   21.67$\pm$0.27      \\
   $D_L$, Mpc$^{*}$        &       217.8$\pm$0.3     \\
 $M_g$, mag$^\dag$         & $-$16.51$\pm$0.06     \\
log $L_g$/L$_{g,\odot}$$^\ddag$&     8.79$\pm$0.44     \\
log $M_\star$/M$_\odot$$^{\dag\dag}$&   6.25$\pm$0.15       \\
$M_\star$/$L_g$, M$_\odot$/L$_{g,\odot}$& 0.0029$\pm$0.0001       \\
$L$(H$\beta$), erg s$^{-1}$$^{**}$&(3.5$\pm$0.6)$\times$10$^{40}$\\
SFR, M$_\odot$yr$^{-1}$$^{\ddag\ddag}$  &     0.77$\pm$0.13 \\
sSFR, Gyr$^{-1}$          &    430$\pm$180 \\
  12+logO/H$^{\dag\dag\dag}$              &7.082$\pm$0.016 \\
\hline
  \end{tabular}


\noindent$^{*}$Luminosity distance.

\noindent$^\dag$Absolute magnitude corrected for Milky Way extinction.

\noindent$^\ddag$log of the SDSS $g$-band luminosity corrected for Milky Way extinction.

\noindent$^{\dag\dag}$Stellar mass derived from the extinction-corrected SDSS spectrum.

\noindent$^{**}$H$\beta$ luminosity derived from the extinction-corrected SDSS 
spectrum.

\noindent$^{\ddag\ddag}$Star formation rate derived from the \citet{K98} relation
 using the extinction-corrected H$\beta$ luminosity.

\noindent$^{\dag\dag\dag}$Oxygen abundance derived from the SDSS spectrum.

  \end{table}

\begin{figure*}
\hbox{
\includegraphics[angle=-90,width=0.85\linewidth]{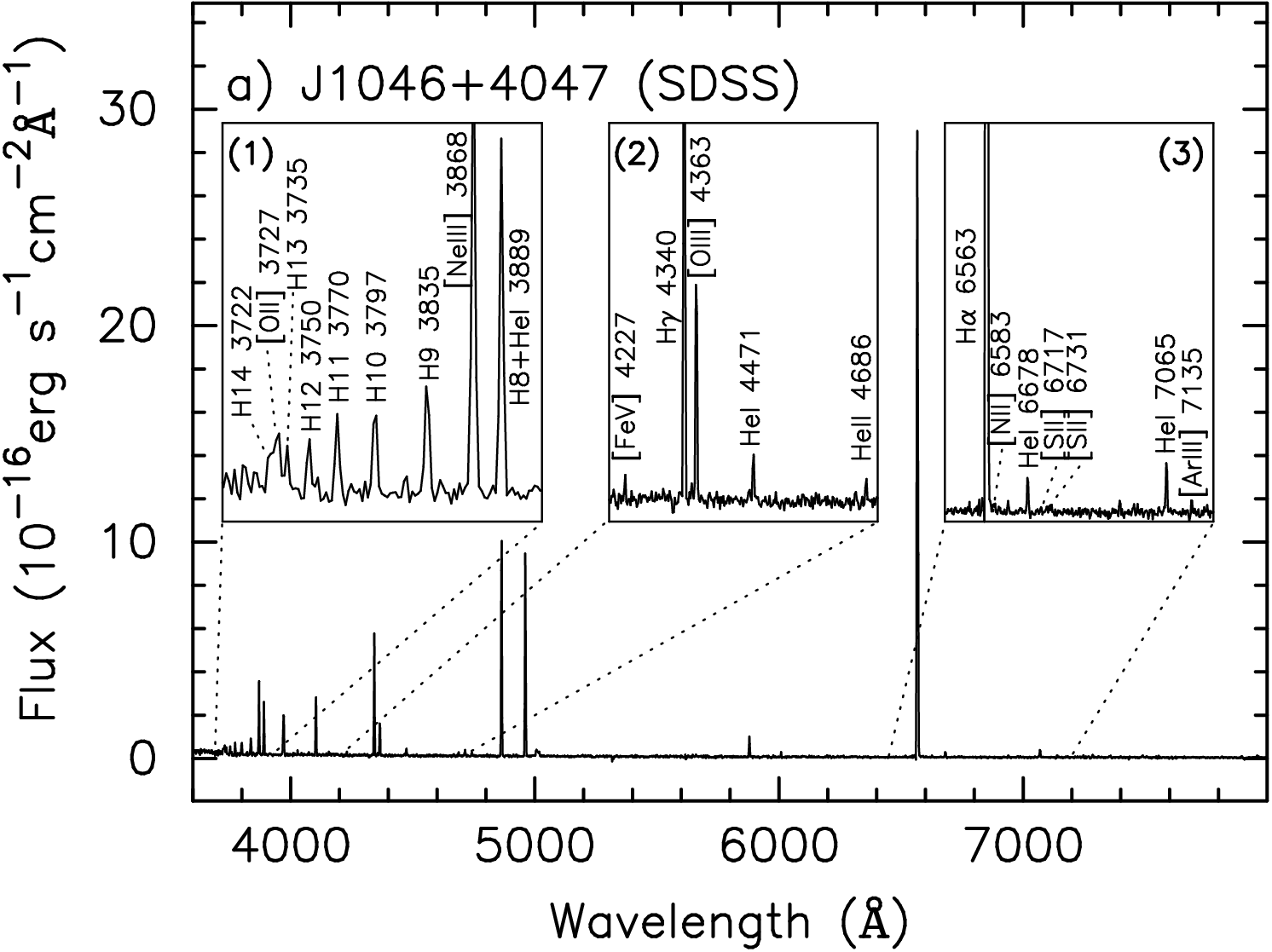}
}
\vspace{0.2cm}
\hbox{
\includegraphics[angle=-90,width=0.85\linewidth]{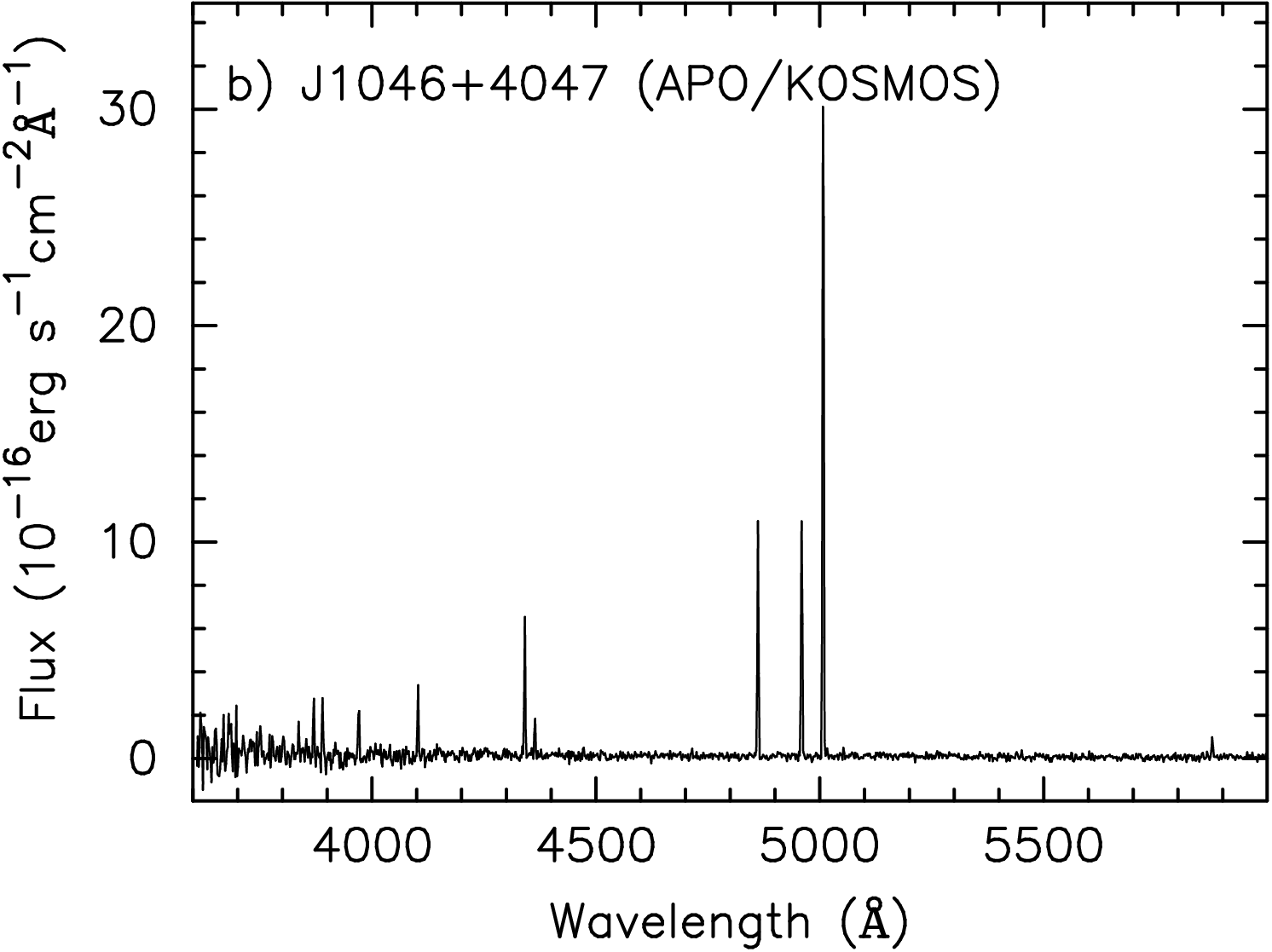}
}
\caption{The rest-frame (a) SDSS spectrum and (b) APO/KOSMOS
spectrum of J1046$+$4047.
Three insets in (a) show expanded parts of the spectral regions for a
better view of weak features. Some emission lines are labelled. 
}
\label{fig1}
\end{figure*}

\section{Observations and data reduction}\label{sec:observations}

\subsection{SDSS and APO/KOSMOS spectra}

We show in Fig.~\ref{fig1}a the SDSS spectrum of J1046$+$4047. It is seen from
insets (1) -- (3) that the signal-to-noise ratio of most weak emission lines is
sufficiently high to derive extinction from the hydrogen Balmer emission
lines and the element abundances. In particular, [O~{\sc iii}] $\lambda$4363
emission line [inset (2)] is strong. Another feature is the weak
[O~{\sc ii}] $\lambda$3727 emission line which is partially blended with
the hydrogen H13 and H14 emission lines. To
derive its intensity, we apply the deblending procedure with fitting by
Gaussians. However, the strongest emission lines H$\beta$,
[O~{\sc iii}] $\lambda$4959 and most notably the 
[O~{\sc iii}] $\lambda$5007 emission lines, are clipped.
The [O~{\sc iii}] $\lambda$4959 emission line is skewed because of
clipping and the [O~{\sc iii}] $\lambda$4959/H$\beta$ flux ratio is equal to 0.89 in the
SDSS spectrum, whereas it is 1.06 in the KOSMOS spectrum.
We also 
can not exclude clipping of the H$\beta$ emission line, because its
intensity is similar to that of the [O~{\sc iii}] $\lambda$4959 emission line.
Therefore, the SDSS spectrum alone can not be used for the determination of the
electron temperature and element abundances.

We attempted to observe this galaxy with the 8.4m Large Binocular
Telescope (LBT) to obtain a higher signal-to-noise spectrum and to derive
intensities of the strong emission lines. However, due to a specific particularity
of the Multi-Object Double Spectrograph (MODS) at the LBT, it was not possible
to select a guide star for the observations. Therefore, we observed the blue part
of the J1046$+$4047 spectrum with the KOSMOS at the 3.5m APO telescope on
19 April 2023, with an exposure time of 900 s and a resolving power of
$\sim$2600. Observations were done at the relatively high airmass of 1.36, 
but with the
slit oriented along the parallactic angle. Therefore, the effect of differential
refraction was minimal \citep{F82}, at least in the wavelength range of
$\sim$ 4300 -- 5000\AA, which is of main interest here. The seeing
during the observations was 1.2 arcsec.

Our main aim is to obtain the intensities of the strong emission lines
in the KOSMOS spectrum. This spectrum is shown
in Fig.~\ref{fig1}b. It has a lower signal-to-noise ratio compared to the SDSS
spectrum. Furthermore, the sensitivity of KOSMOS at $\lambda$~$<$~4000\AA\ is low,
making uncertain the determination of line intensities at these wavelengths,
including the [O~{\sc ii}]~$\lambda$3727 emission line. Therefore, for 
element abundance determination, we use the SDSS spectrum. The intensities of
H$\beta$, [O~{\sc iii}] $\lambda$4959, 5007 emission lines in this spectrum
are scaled from the KOSMOS spectrum with the use of the H$\gamma$ emission line
intensities in both the SDSS and KOSMOS spectra. The observed intensities
of the H$\gamma$ emission line in the SDSS and KOSMOS spectra differ only by
$\sim$ 5 per cent, indicating that aperture corrections for 
different spectroscopic apertures are small. We also note that
the intensity of the H$\beta$ emission line in the SDSS spectrum is
$\sim$ 10 per cent lower than in the KOSMOS spectrum and its equivalent width
in the SDSS spectrum of $\sim$ 390\AA\ is lower than the one of $\sim$ 411\AA\ in the
KOSMOS spectrum. These differences indicate some small clipping of the H$\beta$
emission line in the SDSS spectrum.

The observed emission-line intensities in both
SDSS and KOSMOS spectra were measured using the {\sc iraf} {\it splot} routine.
The emission-line errors in the SDSS spectrum are calculated using the
pixel-by-pixel errors included in the file of the spectrum, whereas the
emission-line errors in the KOSMOS spectrum are calculated using the
pixel-by-pixel errors generated by the {\sc iraf} {\it apall} routine during 
extraction of the one-dimensional spectrum.
We then use the ratio of the H$\gamma$ emission line intensities in both
SDSS and KOSMOS spectra to scale the intensities of the strong H$\beta$,
[O~{\sc iii}] $\lambda$4959, 5007 emission lines in the KOSMOS spectrum.
These intensities are employed instead of those of the clipped emission
lines in the SDSS spectrum.

The intensities of emission lines in the
SDSS spectrum were corrected iteratively for extinction and underlying 
stellar absorption, derived from comparison of the observed intensity
ratios of the hydrogen Balmer emission lines H$\alpha$, H$\beta$,
H$\gamma$, 
H$\delta$, H9, H10, H11, H12 with the theoretical recombination ratios,
as described in \citet{ITL94}. The theoretical ratios weakly depend
on the electron temperature and electron number density.
We obtain them, adopting the approximate
electron temperature $T_{\rm e}$(O~{\sc iii}) from the observed
[O~{\sc iii}]~$\lambda$4363/($\lambda$4959+$\lambda$5007) intensity ratio, and
the approximate electron number density from the observed
[S~{\sc ii}]~$\lambda$6717/$\lambda$6731 intensity ratio.
The equivalent widths of the underlying stellar Balmer 
absorption lines are assumed to be the same for each hydrogen line.

The extinction-corrected fluxes together with the extinction 
coefficient $C$(H$\beta$), the observed H$\beta$ emission-line flux 
$F$(H$\beta$), the rest-frame equivalent width EW(H$\beta$) of the H$\beta$ 
emission line, and the equivalent width of the Balmer absorption lines
are shown in the left column of Table~\ref{tab2}. 

\subsection{Enhancement of H$\alpha$ emission}

We find the derived extinction coefficient $C$(H$\beta$) to be abnormally high,
equal to 0.520. This value is
not typical for low-metallicity SFGs. Furthermore, the
extinction-corrected intensities of H$\gamma$ and other higher-order
hydrogen lines in the
blue part of the spectrum relative to the extinction-corrected intensity of
the H$\beta$ line are $\sim$ 20 -- 30 per cent higher than the theoretical
recombination values \citep[e.g. ][]{SH95}. It is clear that
these emission-line intensities were overcorrected for extinction. The likely reason for
such an overcorrection may be that the H$\alpha$ emission line is enhanced by some
non-recombination process. For example, an additional contribution of
H$\alpha$ emission from massive stars and supernovae could produce such an effect. Such
an enhancement has been detected in the spectra of several other SFGs, most
notably in the galaxies J1320$+$2155 \citep{IT09}, Haro 11C \citep{G12}, and
J1154$+$2443 \citep{I18b}.

It is clear from the above considerations that the H$\alpha$ emission line
should not be used for the determination of the extinction coefficient in
J1046$+$4047. Excluding this line, we derive a much lower $C$(H$\beta$) = 0.180,
more typical of low-metallicity SFGs.
The intensities of all hydrogen emission lines 
relative to H$\beta$, corrected for extinction with the lower $C$(H$\beta$)
and underlying absorption (right column in Table~\ref{tab2})
are now close to the theoretical recombination values \citep{SH95}.

\subsection{A very young starburst and hard ionizing radiation}

We note two features in the spectrum of J1046$+$4047, which are very rarely
seen
in the spectra of other SFGs, including those with the lowest heavy element 
abundances. First, the O$_{32}$ ratio of 57 is very high compared to the ones in
most other SFGs from the SDSS \citep[compare the position of
another SFG, J2229+2725, with a similar O$_{32}$~=~53, and the positions of SDSS
galaxies in fig.~4 from][]{Iz21}. Second, the 
H$\beta$ equivalent width EW(H$\beta$) of $\sim$~411\AA\ is also very large.
J1046$+$4047 is very similar in these characteristics to
J2229$+$2725 \citep{Iz21}. This indicates that the starbursts in both galaxies
are very young, with an age $\la$~1~--~2~Myr.

The SDSS spectrum of J1046$+$4047 has a sufficiently high signal-to-noise
ratio to show two high-ionization emission lines [Fe~{\sc v}] $\lambda$4227
and He~{\sc ii} $\lambda$4686 [inset (2) in Fig.~\ref{fig1}a and
Table~\ref{tab2}], indicative of the presence of hard ionising radiation (with energies greater than 54 eV). The [Fe~{\sc v}]~$\lambda$4227 emission
line is frequently seen in spectra with high signal-to-noise ratio of
low-metallicity star-forming galaxies \citep*[e.g. ][]{F01,I01,TI05,Iz21b,B21}. Its presence  
is associated with that of the He~{\sc ii} $\lambda$4686 emission line.
On the other hand, the low-ionization lines [N~{\sc ii}] $\lambda$6583 and
[S~{\sc ii}] $\lambda$6717, 6731 are very weak [inset (3) in Fig.~\ref{fig1}a],
indicative of a very high ionization parameter.

\begin{table}
\caption{Extinction-corrected emission-line intensity ratios in J1046$+$4047 \label{tab2}}
\begin{tabular}{lrr} \hline
Line& \multicolumn{2}{c}{100$\times$$I$($\lambda$)/$I$(H$\beta$)} \\ \cline{2-3}
    & $C$(H$\beta$) is derived&$C$(H$\beta$) is derived   \\ 
    &including H$\alpha$& excluding H$\alpha$ \\ \hline
                                &     (1)~~~~~  &     (2)~~~~~  \\ \hline
3722.00 H14                     &  3.44$\pm$1.05&  2.67$\pm$0.81\\
3727.00 [O {\sc ii}]            &  7.20$\pm$1.22&  5.61$\pm$0.95\\
3735.00 H13                     &  4.37$\pm$1.38&  2.85$\pm$0.82\\
3750.15 H12                     &  5.22$\pm$1.27&  3.56$\pm$0.83\\
3770.63 H11                     &  7.85$\pm$1.25&  5.72$\pm$0.89\\
3797.90 H10                     &  8.39$\pm$1.28&  6.15$\pm$0.92\\
3835.39 H9                      & 11.11$\pm$1.37&  8.33$\pm$1.01\\
3868.76 [Ne {\sc iii}]          & 35.62$\pm$2.07& 28.43$\pm$1.65\\
3889.00 He {\sc i}+H8           & 27.52$\pm$1.84& 21.62$\pm$1.44\\
3968.00 [Ne {\sc iii}]+H7       & 33.24$\pm$1.99& 26.60$\pm$1.58\\
4026.19 He {\sc i}              &  1.89$\pm$0.84&  1.56$\pm$0.69\\
4101.74 H$\delta$               & 35.02$\pm$2.00& 28.87$\pm$1.64\\
4227.20 [Fe {\sc v}]            &  1.50$\pm$0.74&  1.29$\pm$0.63\\
4340.47 H$\gamma$               & 54.23$\pm$2.38& 47.71$\pm$2.09\\
4363.21 [O {\sc iii}]           & 17.12$\pm$1.32& 15.27$\pm$1.17\\
4471.48 He {\sc i}              &  3.69$\pm$0.79&  3.38$\pm$0.73\\
4685.94 He {\sc ii}             &  1.65$\pm$0.59&  1.59$\pm$0.57\\
4712.00 [Ar {\sc iv}]+He {\sc i}&  2.95$\pm$0.66&  2.86$\pm$0.64\\
4740.20 [Ar {\sc iv}]           &  1.94$\pm$0.59&  1.89$\pm$0.58\\
4861.33 H$\beta$                &100.00$\pm$3.22&100.00$\pm$3.21\\
4958.92 [O {\sc iii}]           &103.92$\pm$3.27&106.30$\pm$3.33\\
5006.80 [O {\sc iii}]           &309.65$\pm$7.55&319.76$\pm$7.77\\
5875.60 He {\sc i}              &  7.76$\pm$0.69&  9.12$\pm$0.81\\
6562.80 H$\alpha$               &281.88$\pm$7.12&356.59$\pm$8.98\\
6583.40 [N~{\sc ii}]            &  $<$0.41~~~~~&  $<$0.53~~~~~\\
6678.10 He {\sc i}              &  2.04$\pm$0.36&  2.61$\pm$0.46\\
6716.40 [S~{\sc ii}]            &  0.34$\pm$0.24&  0.44$\pm$0.30\\
6730.80 [S~{\sc ii}]            &  0.36:$\pm$0.24&  0.46:$\pm$0.31\\
7065.30 He {\sc i}              &  2.92$\pm$0.39&  3.90$\pm$0.52\\
7135.80 [Ar~{\sc iii}]          &  0.64$\pm$0.24&  0.86$\pm$0.32\\
7281.35 He {\sc i}              &  0.86$\pm$0.25&  1.18$\pm$0.35\\ \\
$C$(H$\beta$)$^{\dag}$        &\multicolumn{1}{c}{0.520$\pm$0.030}&\multicolumn{1}{c}{0.180$\pm$0.030}\\
$F$(H$\beta$)$^{\ddag}$        &\multicolumn{1}{c}{41.13$\pm$2.25}&\multicolumn{1}{c}{41.13$\pm$2.25}\\
EW(H$\beta$)$^{*}$           &\multicolumn{1}{c}{411.3$\pm$22.0}&\multicolumn{1}{c}{411.3$\pm$22.0}\\
EW(abs)$^{*}$                &\multicolumn{1}{c}{1.1$\pm$0.3}&\multicolumn{1}{c}{0.0$\pm$0.0}\\
\hline
  \end{tabular}






\hbox{$^{\dag}$Extinction coefficient, derived from the observed hydrogen} 

\hbox{\,~Balmer decrement.}

\hbox{$^{\ddag}$Observed flux in units of 10$^{-16}$ erg s$^{-1}$ cm$^{-2}$.}

\hbox{$^{*}$Rest-frame equivalent width in \AA.}

  \end{table}

\begin{figure}
\hbox{
\includegraphics[angle=-90,width=0.81\linewidth]{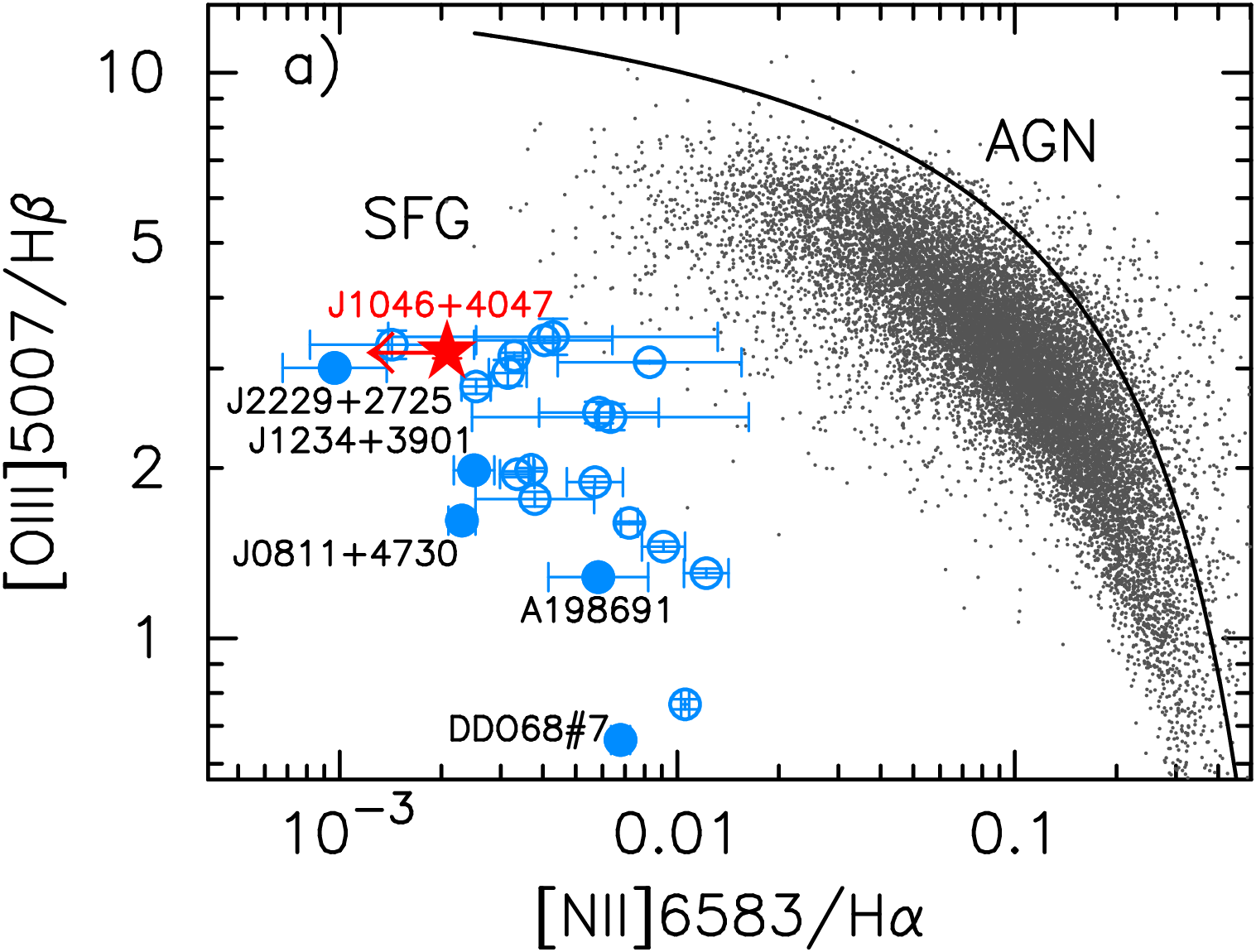}
}
\vspace{0.2cm}
\hbox{
\includegraphics[angle=-90,width=0.81\linewidth]{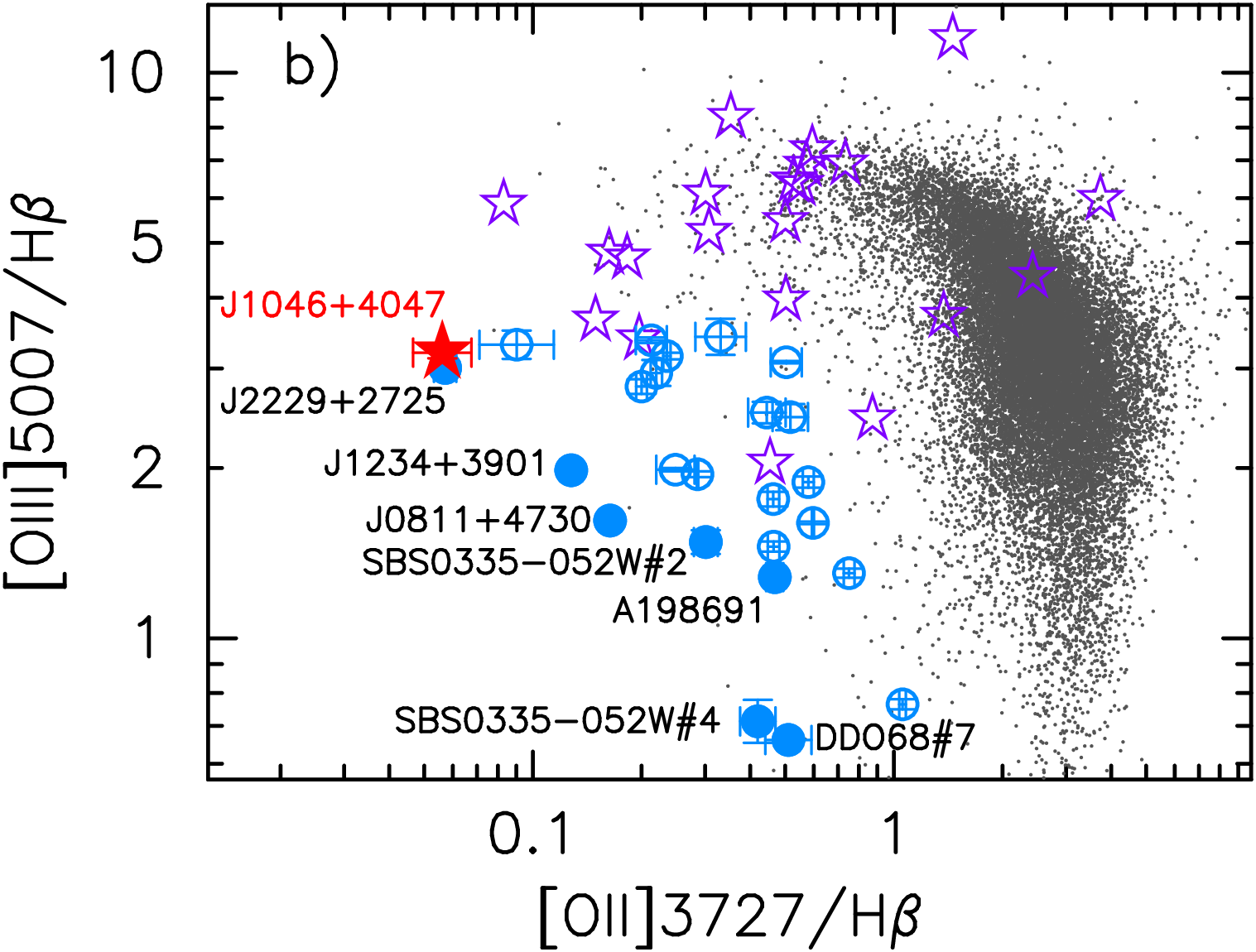}
}
\vspace{0.2cm}
\hbox{
\includegraphics[angle=-90,width=0.81\linewidth]{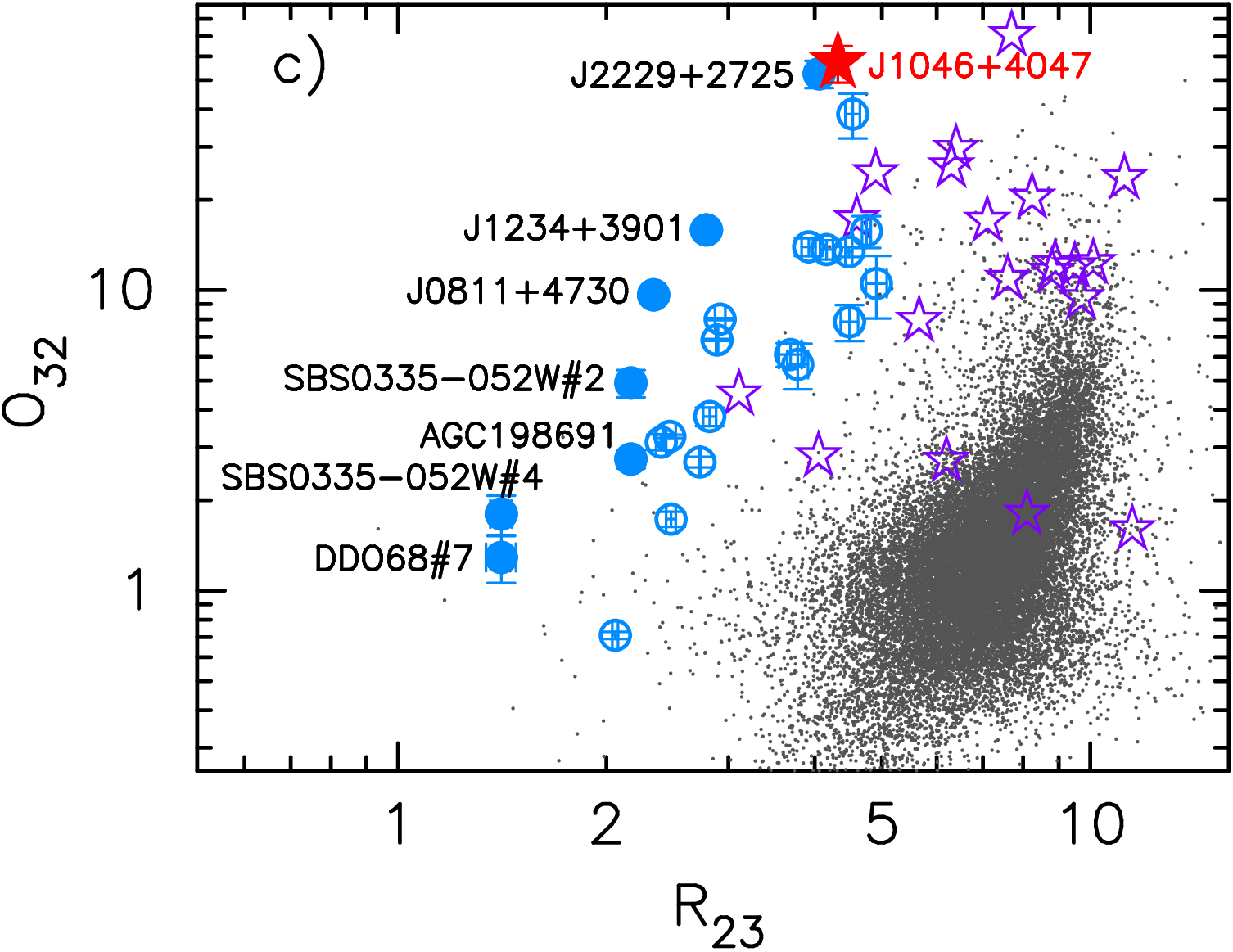}
}
\caption{(a) The Baldwin-Phillips-Terlevich (BPT) diagram
$I$([O~{\sc iii}]$\lambda$5007)/$I$(H$\beta$) -- $I$([N~{\sc ii}]$\lambda$6563)/$I$(H$\alpha$) \citep*{BPT81};
(b) the $I$([O~{\sc iii}]$\lambda$5007)/$I$(H$\beta$) -- 
$I$([O~{\sc ii}]$\lambda$3727)/$I$(H$\beta$) diagram;
(c) the O$_{32}$ -- R$_{23}$ diagram for SFGs, where O$_{32}$= 
$I$([O~{\sc iii}]$\lambda$5007)/$I$([O~{\sc ii}]$\lambda$3727) and R$_{23}$ = 
$I$([O~{\sc ii}]$\lambda$3727 + [O~{\sc iii}]$\lambda$4959 + 
[O~{\sc iii}]$\lambda$5007)/$I$(H$\beta$). The lowest-metallicity SFGs
and H~{\sc ii} regions in SFGs with 12 + log(O/H) $<$ 7.1
from \citet[][H~{\sc ii} region DDO68\#7]{An19},
\citet[][H~{\sc ii} regions SBS0335$-$052W\#4 and SBS0335$-$052W\#2]{I09},
\citet[][compact SFG A198691]{H16},
\citet[][compact SFG J0811$+$4730]{I18a},
\citet[][compact SFG J1234$+$3901]{Iz19a}, 
\citet[][compact SFG J2229$+$2725]{Iz21} are shown by labelled blue filled
circles.
The SFGs with 12 + log(O/H) in the range 7.1 -- 7.3 from \citet{Iz21} are
represented by open blue circles.
The location of compact SFG J1046$+$4047 adopting $C$(H$\beta$) = 0.180
(Table~\ref{tab2}) is shown by a filled star.
High-$z$ SFGs with $z$ $\sim$ 6 -- 8 \citep{He22,Sa23} are represented
in (b) and (c) by open purple stars and SFGs from the SDSS DR16 are
shown in all panels by grey dots.}
\label{fig2}
\end{figure}

\section{Diagnostic diagrams and integrated characteristics of J1046$+$4047}
\label{sec:integr}

\subsection{Diagnostic diagrams}

It is known \citep*[e.g. ][]{I12,H16,I18a} that extremely metal-deficient
SFGs strongly deviate in the diagnostic diagrams from the main
sequence SFGs with higher metallicities. This difference can be used to search
for the most metal-deficient galaxies by analyzing the relations between
the strongest emission lines in the optical spectra.

The most commonly used diagnostic diagram is the one proposed by Baldwin, Phillips \& Terlevich (BPT)
\citep{BPT81}. It is shown in Fig.~\ref{fig2}a. The most metal-deficient SFGs
with 12 + log(O/H) $<$ 7.1 (filled circles) and the SFGs with
12 + log(O/H) $=$ 7.1 -- 7.3 (open circles) are located far from the
main-sequence SFGs (grey dots). The latter are in general much more enriched
with heavy elements, having a characteristic 12 + log(O/H) in the range
$\sim$ 7.9 -- 8.1. The SFGs shown by coloured symbols form two well-separated distinct sequences: SFGs shown by filled circles have lower
metallicities compared to SFGs represented by open circles. The
[N~{\sc ii}] $\lambda$6583 emission line is not detected in the spectrum
of J1046$\pm$4047. Therefore we show its location in Fig.~\ref{fig2}a
(filled star) by plotting the upper limit of this line. Furthermore, all extremely
low-metallicity SFGs are faint objects with a very weak
[N~{\sc ii}] $\lambda$6583 emission line, so that it can be detected only in
spectra with high signal-to-noise ratios.

A more promising diagram to search for the most metal-poor SFGs is the
$I$([O~{\sc iii}]~$\lambda$5007)/$I$(H$\beta$) -- $I$([O~{\sc ii}]~$\lambda$3727)/$I$(H$\beta$) diagram (Fig.~\ref{fig2}b), because it relies on the
considerably brighter [O~{\sc ii}]~$\lambda$3727 emission line rather than the
weaker [N~{\sc ii}]~$\lambda$6583
emission line. Similarly to Fig.~\ref{fig2}a, the lowest-metallicity SFGs form
two well-separated sequences. Finally, we consider the O$_{32}$~--~R$_{23}$ diagram,
shown in Fig.~\ref{fig2}c, where O$_{32}$= 
$I$([O~{\sc iii}]$\lambda$5007)/$I$([O~{\sc ii}]$\lambda$3727) and R$_{23}$ = 
$I$([O~{\sc ii}]$\lambda$3727 + [O~{\sc iii}]$\lambda$4959 + 
[O~{\sc iii}]$\lambda$5007)/$I$(H$\beta$). Again, two distinct sequences of
lowest-metallicity SFGs are seen. We note that the locations of
two SFGs, J1046$+$4047 and J2229$+$2725 are nearly coincident in
Figs.~\ref{fig2}b and \ref{fig2}c, indicating that both galaxies have nearly the
same oxygen abundances. For comparison, the $z$ $\sim$ 6 -- 8 SFGs
of \citet{He22} and \citet{Sa23} are shown in Fig.~\ref{fig2}b and
Fig.~\ref{fig2}c by open purple stars. The
positions of most of these SFGs indicate a metallicity lower than those of 
main-sequence SFGs, but not as low as those in the nearby most metal-deficient
galaxies with 12~+~log(O/H)~$<$~7.1.

\subsection{Integrated characteristics}

For the determination of the integrated characteristics of J1046$+$4047
such as stellar mass, star-formation rate and luminosity, we have adopted
the luminosity distance
$D_L$ = 347 Mpc, obtained from the galaxy redshift for the cosmological 
parameters H$_0$ = 67.1 km s$^{-1}$ Mpc$^{-1}$, $\Omega_m$ = 0.318, 
$\Omega_\Lambda$ = 0.682 \citep{P14}.

The stellar mass of J1046$+$4047 is determined from fitting the 
spectral energy distribution (SED). We follow the prescriptions of
\citet{I18a}, {\sc starburst99} models \citep{L99} and adopt stellar evolution
models by \citet{G00}, stellar atmosphere models by \citet*{L97} and the
\citet{S55} initial mass function (IMF) with lower and upper mass limits of
0.1 M$_\odot$ and 100 M$_\odot$, respectively.

The equivalent width of the H$\beta$ emission line in J1046$+$4047 of 411\AA\
is high (Table \ref{tab2}), indicating a substantial contribution 
of the nebular continuum, amounting about 50 per cent of the total continuum
near the H$\beta$ emission line. Therefore, following e.g. \citet{I18a}, both 
the stellar and nebular continua are taken into account in the SED fitting.
The star-formation history was approximated by a recent short burst at age
$t_{\rm b}$ $<$ 10 Myr and a prior continuous star formation.
Details on the SED fitting procedure can be found e.g. in \citet{I18a}.

We obtain a stellar mass $M_\star$ = 10$^{6.25}$~M$_\odot$ for
J1046$+$4047. The star formation rate of 0.77 M$_\odot$\,yr$^{-1}$ is
derived from the \citet{K98}
relation using the extinction-corrected H$\beta$ luminosity and the
H$\alpha$/H$\beta$ recombination intensity ratio of 2.7.
This yields a very high specific star formation rate sSFR of 
$\sim$ 430 Gyr$^{-1}$, indicative of very active ongoing star formation.
The absolute SDSS $g$ magnitude, corrected for the Milky Way 
extinction is $M_g$ = $-$16.51 mag (Table \ref{tab1}). 
Adopting $M_\star$ = 10$^{6.25}$~M$_\odot$, we derive the very low 
mass-to-luminosity ratio $M_\star$/$L_g$ of 0.0029
(in solar units) for J1046$+$4047. That ratio is comparable to those of three other most
metal-poor compact SFGs, J0811$+$4730, J1234$+$3901 and J2229$+$2725
\citep{I18a,Iz19a,Iz21}. It is approximately two orders of magnitude lower than
the mass-to-light ratios of the majority of compact SFGs in the SDSS
shown by grey dots in Fig.~\ref{fig2}.

\begin{table}
\caption{Electron temperatures, electron number density 
and heavy element abundances in J1046$+$4047 \label{tab3}}
\begin{tabular}{lcc} \hline
Property    & \multicolumn{2}{c}{Value} \\ \cline{2-3}
    & $C$(H$\beta$) is derived&$C$(H$\beta$) is derived   \\ 
&including H$\alpha$& excluding H$\alpha$ \\ \hline
                                    &       (1)       &       (2)            \\ \hline
$T_{\rm e}$(O {\sc iii}), K           &  28300$\pm$2000 & 25300$\pm$1600       \\
$T_{\rm e}$(O {\sc ii}), K           &   22800$\pm$800 &   20700$\pm$700       \\
$T_{\rm e}$(S {\sc iii}), K          &   25200$\pm$1700&   22700$\pm$1300       \\
$N_{\rm e}$(S {\sc ii}), cm$^{-3}$    &920$\pm$1800     &880$\pm$1800         \\ \\
O$^+$/H$^+$$\times$10$^6$            &0.232$\pm$0.034 &0.227$\pm$0.033 \\
O$^{2+}$/H$^+$$\times$10$^5$          &0.931$\pm$0.045 &1.163$\pm$0.043 \\
O$^{3+}$/H$^+$$\times$10$^6$          &0.224$\pm$0.091 &0.212$\pm$0.086 \\
O/H$\times$10$^5$                   &0.977$\pm$0.046 &1.207$\pm$0.044 \\
12+log(O/H)                         &6.990$\pm$0.021 &7.082$\pm$0.016     \\ \\
N$^{+}$/H$^+$$\times$10$^7$          &$<$0.208 &$<$0.307 \\
ICF(N)                              &36.672 &46.133 \\
N/H$\times$10$^6$                   &$<$0.763 &$<$1.417 \\
log(N/O)                            &$<$$-$1.108&$<$$-$0.931 \\ \\
Ne$^{2+}$/H$^+$$\times$10$^6$        &2.220$\pm$0.132 &2.184$\pm$0.127 \\
ICF(Ne)                             &1.021 &1.017 \\
Ne/H$\times$10$^6$                  &2.267$\pm$0.135 &2.221$\pm$0.130 \\
log(Ne/O)                           &$-$0.634$\pm$0.033~ &$-$0.735$\pm$0.030~\\ \\
Ar$^{2+}$/H$^+$$\times$10$^7$        &0.120$\pm$0.058 &0.185$\pm$0.091 \\
ICF(Ar)                             &2.993 &3.557 \\
Ar/H$\times$10$^7$                  &0.359$\pm$0.504&0.658$\pm$0.748 \\
log(Ar/O)                           &$-$2.435$\pm$0.610~&$-$2.263$\pm$0.494~ \\ \\
\hline
\end{tabular}





  \end{table}

\section{Heavy element abundances}\label{sec:abundances}

\subsection{Direct method}

We use the extinction-corrected emission-line fluxes derived from the SDSS
spectrum and the prescriptions of
\citet{I06} to derive heavy element abundances in 
J1046$+$4047. The electron temperature $T_{\rm e}$(O~{\sc iii}) is calculated 
from the [O~{\sc iii}]$\lambda$4363/($\lambda$4959 + $\lambda$5007) 
emission-line flux ratio. It is used to derive the abundances of O$^{3+}$, 
O$^{2+}$ and Ne$^{2+}$.
The abundances of O$^{+}$, N$^{+}$ and S$^{+}$ are derived with the electron
temperature $T_{\rm e}$(O~{\sc ii}) and the abundance of Ar$^{2+}$
is derived with the electron temperature $T_{\rm e}$(S~{\sc iii}), using the 
relations of \citet{G92} between $T_{\rm e}$(O~{\sc ii}), $T_{\rm e}$(S~{\sc iii})
and $T_{\rm e}$(O~{\sc iii}). The electron number density was derived from the
[S~{\sc ii}]$\lambda$6717/$\lambda$6731 flux ratio. The electron temperatures
and electron number densities are shown in Table~\ref{tab3}. We note that the
[S~{\sc ii}]$\lambda$6717, $\lambda$6731 emission lines are very weak
[inset (3) in Fig.~\ref{fig1}a]. Therefore, the  electron number density is
uncertain.

For comparison, we first consider the case with $C$(H$\beta$) = 0.520, which is the value derived when
the H$\alpha$ emission line is included in the determination of extinction. The
emission-line intensities are shown in column (1) of Table~\ref{tab2}.
The derived electron temperature $T_{\rm e}$(O~{\sc iii}) of 28300 K 
is very high (Table~\ref{tab3}). It is likely to be overestimated because of the
overestimation of $C$(H$\beta$). The electron number density is also higher
than typical values for SFGs, though it is very uncertain because of
the weakness of [S~{\sc ii}]$\lambda$6717, $\lambda$6731 emission lines
[inset (3) in Fig.~\ref{fig1}a]. We note that uncertainties in the electron
number density play a minor role as long as the low-density limit is
satisfied, i.e. when the observed electron number density is less than critical
densities of $\ga$ 10$^{5}$ cm$^{-3}$ for forbidden transitions, including
[S~{\sc ii}]$\lambda$6717, $\lambda$6731 transitions. That is the case here.
The ionic abundances, ionisation correction factors and total O, N, Ne and 
Ar abundances are obtained using relations by \citet{I06}. They are 
presented in Table~\ref{tab3}. We derived the extremely low oxygen abundance
12 + log(O/H) of 6.99, in part due to an overestimated $T_{\rm e}$(O~{\sc iii}).
The logarithm of the Ne/O abundance ratio log(Ne/O) = --0.634 is somewhat
higher compared
to the values of $-$0.75 -- $-$0.85 for SFGs with low metallicities
\citep[e.g. ][]{I06}. This is because the [Ne~{\sc iii}]~$\lambda$3868
emission line, used for the Ne abundance determination, has been overcorrected for
extinction as compared to the [O~{\sc iii}] emission lines.

Adopting the more reasonable value $C$(H$\beta$) = 0.180, derived by excluding the H$\alpha$
emission line, we obtain $T_{\rm e}$(O~{\sc iii}) = 25300 K [column (2) in
Table~\ref{tab3}]. Although this temperature is high, it is still similar to that in other
most metal-deficient SFGs.
The nebular oxygen abundance of 12+log(O/H) = 7.082$\pm$0.016 in J1046$+$4047
is very close to that in J2229$+$2725 \citep{Iz21}. It is among the lowest
found for SFGs. The logarithms of the Ne/O and Ar/O abundance ratios
(Table~\ref{tab3}) for this galaxy are similar to those in other
low-metallicity SFGs \citep[e.g., ][]{I06}.

\begin{figure}
\hbox{
\includegraphics[angle=-90,width=0.99\linewidth]{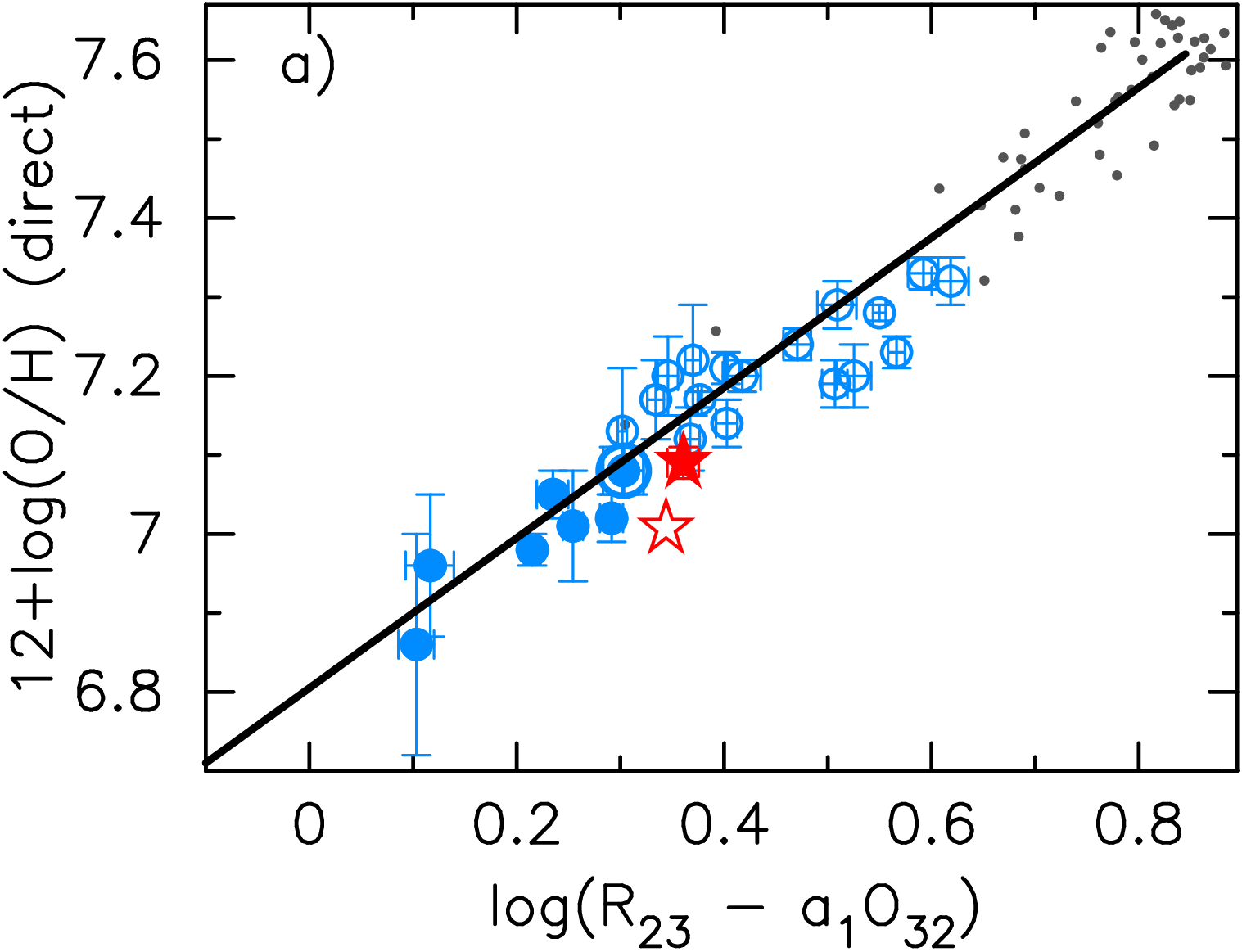}
}
\vspace{0.2cm}
\hbox{
\includegraphics[angle=-90,width=0.99\linewidth]{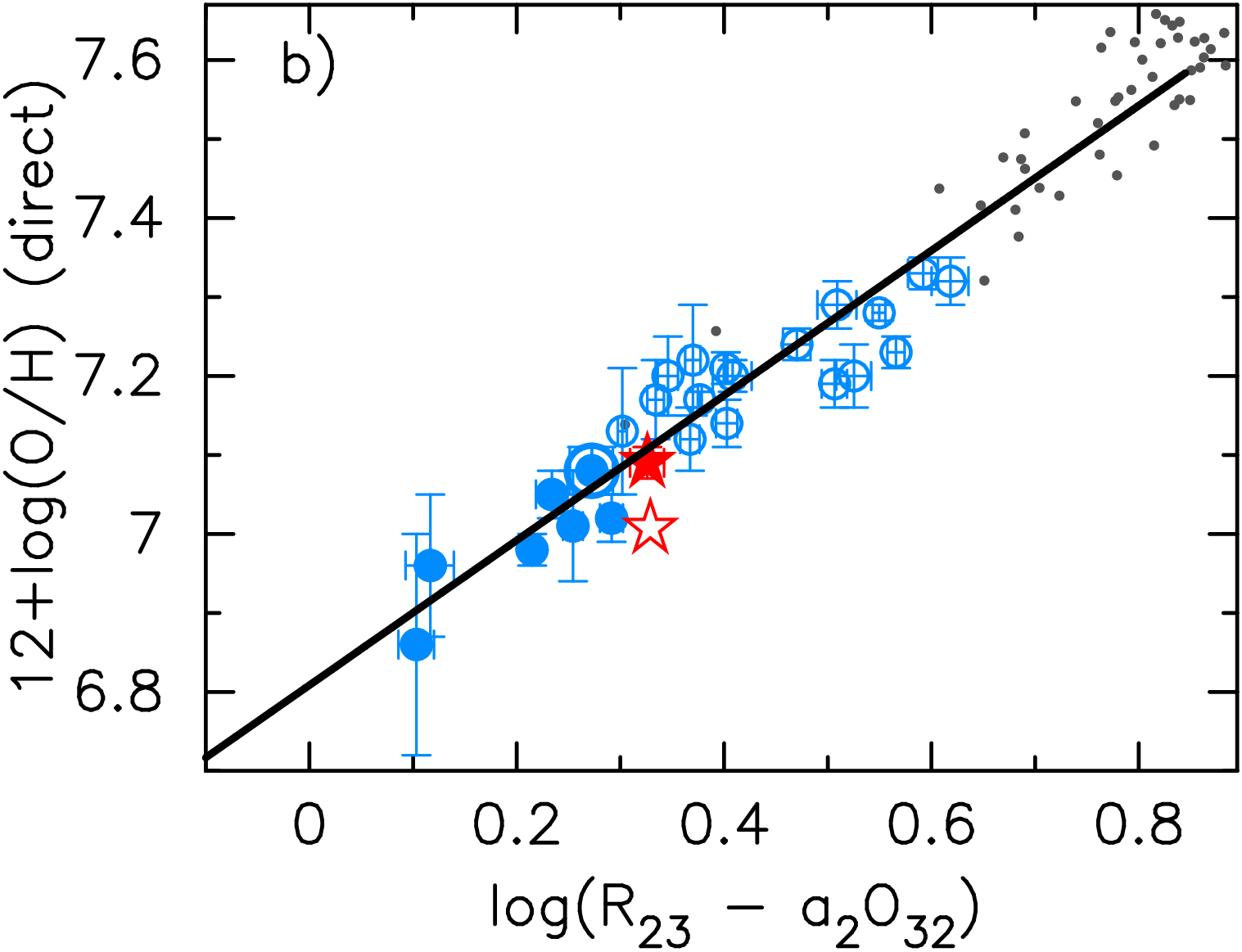}
}
\caption{The (a) log(R$_{23}$ - $a_1$O$_{32}$) - (12 + log(O/H)), and
(b) log(R$_{23}$ - $a_2$O$_{32}$) - (12 + log(O/H)) relations, where $a_1$ = 0.080 -- 0.00078O$_{32}$ \citep{Iz21} and $a_2$ = $a_1$ + 0.00000095O$_{32}^2$.
In both relations 12 + log(O/H) is derived by the direct method, R$_{23}$ = 
($I$([O~{\sc iii}]$\lambda$4959+$\lambda$5007 + [O~{\sc ii}]$\lambda$3727))/$I$(H$\beta$) and 
O$_{32}$ = $I$([O~{\sc iii}]$\lambda$5007)/$I$([O~{\sc ii}]$\lambda$3727).
The galaxy J1046$+$4047 (this paper) is represented by a 
red filled star for the extinction coefficient $C$(H$\beta$) = 0.180 and by a 
red open star for the extinction coefficient $C$(H$\beta$) =
0.520 (Table~\ref{tab2}). Other lowest-metallicity SFGs from \citet{Iz19b,Iz21}
are shown by blue filled circles (12 + log(O/H) $<$ 7.1) and blue open circles
(12 + log(O/H) = 7.1 - 7.3). The galaxy J2229$+$2725  \citep{Iz21} with
O$_{32}$ = 53 is encircled. SFGs from the SDSS with 12 + log(O/H) $\la$ 7.65 and
an [O~{\sc iii}] $\lambda$4363 emission line measured with a signal-to-noise ratio
$\geq$ 10 are represented by grey dots. The solid lines in (a) and (b) are
linear maximum-likelihood fits \citep{Pr07} obtained by \citet{Iz21} and
in this paper, respectively.}
\label{fig3}
\end{figure}


\subsection{Strong-line method for oxygen abundance determination in XMD galaxies}

The extraordinarily high O$_{32}$ ratio of 57 in J1046$+$4047, together with
O$_{32}$ = 53 in J2229$+$2735, makes both objects excellent targets for testing
and improving the calibration of the strong-line method for oxygen abundance
determination in rare extremely low-metallicity SFGs, with the highest O$_{32}$
ratios, using the strong oxygen emission lines in the optical range.

Recently \citet{Iz19b,Iz21}, have developed a strong-line method for XMD galaxies,
calibrated by using high-quality observations of the most
metal-poor galaxies with a well-detected [O~{\sc iii}]$\lambda$4363 emission
line. A common problem of the strong-line methods is that they depend on several
parameters. In particular, the method based on oxygen line intensities 
depends not only on the sum R$_{32}$ = $I$([O~{\sc ii}]$\lambda$3727 + 
[O~{\sc iii}]$\lambda$4959 + [O~{\sc iii}]$\lambda$5007])/$I$(H$\beta$), but
also on the ionisation parameter. A good indicator of the ionisation parameter
is the O$_{32}$ ratio. 
\citet{Iz21} have proposed a method which takes into account the correction for
ionization parameter in a wide range of O$_{32}$, up to the value of 53.
The dependence found by \citet{Iz21} is shown in Fig.~\ref{fig3}a. It is derived
by the maximum likelihood estimation technique \citep{Pr07},
\begin{equation}
12+\log\frac{\rm O}{\rm H} = 0.950 \log({\rm R}_{23} - a_1{\rm O}_{32}) + 6.805,
\label{eq1}
\end{equation}
where $a_1$ = 0.080 -- 0.00078O$_{32}$. In Fig.~\ref{fig3}a we also show the
position of J1046$+$4047 by a filled star, for $C$(H$\beta$) = 0.180, and
by an open star, for $C$(H$\beta$) = 0.520. For
comparison, the position of J2229$+$2725 is shown by an encircled filled circle.
For $C$(H$\beta$) = 0.180, J1046$+$4047 closely follows the relation for SFGs
with 12 + log(O/H) $<$ 7.3, represented by a straight line, whereas a higher
offset is present for the case with $C$(H$\beta$) = 0.520 (open star). 

However,
a slight offset is present even for the case with $C$(H$\beta$) = 0.180
(filled star). This 
can be due to the fact that, at the high O$_{32}$ value for J1046$+$4047, some
improvement of Eq.~\ref{eq1} is needed.  We can achieve this by adding one more
term in the polynomial expansion of the dependence of the oxygen abundance on the O$_{32}$ ratio, and by minimizing the position deviations of all galaxies in
Fig.~\ref{fig3}, including both J2229$+$2725 (encircled
filled circle) and J1046$+$4047 (filled star), from the linear maximum
likelihood fit. We then iteratively vary the coefficient in the new
polynomial term and calculate a new fit in each step with a new value of
the polynomial $a_2$. We obtain:
\begin{equation}
12+\log\frac{\rm O}{\rm H} = 0.917 \log({\rm R}_{23} - a_2{\rm O}_{32}) + 6.804,
\label{eq2}
\end{equation}
where $a_2$ = 0.080 -- 0.00078O$_{32}$ + 0.00000095O$_{32}^2$.
This relation is shown in Fig.~\ref{fig3}b. Adding a new term only 
changes slightly the positions of J2229$+$2725 and J1046$+$4047, making them in
better agreement with the relation shown by the solid line.

On the other hand, the large offset 
of J1046$+$4047 when adopting $C$(H$\beta$) = 0.520 (open star), indicates that
the oxygen abundance derived by the direct method with this high $C$(H$\beta$) is
underestimated by $\sim$ 0.1 dex. As for the position of J2229$+$2725
(encircled filled blue circle), it is changed to a lesser
extent. The positions of other extremely low metallicity galaxies with 
12+log(O/H) $\la$ 7.3 and lower O$_{32}$ are not changed. We also note that
the extrapolation of the relations Eqs.~\ref{eq1}~--~\ref{eq2} to higher
12+log(O/H), up to 7.65, fits well the position of SDSS SFGs (grey dots in
Fig.~\ref{fig3}).
There is an improvement of 12 + log(O/H) by $\sim$ 0.04 dex in J1046$+$4047
if Eq.~\ref{eq2} is used instead of Eq.~\ref{eq1}. On the other
hand, the calibration by \citet{Iz19b} is not applicable to both J1046$+$4047
and J2229$+$2725 despite their very similar high O$_{32}$, as illustrated in
fig.~8 of \citet{Iz21} for J2229$+$2725.

In summary, both relations Eqs.~\ref{eq1} and \ref{eq2} can be used for the
determination of oxygen abundances in extremely low-metallicity galaxies
with 12+log(O/H) smaller than 7.65, with an uncertainty of $\leq$ 0.05 dex
(Fig.~\ref{fig3}), when the direct method can not be applied because of
the weakness of the [O~{\sc iii}] $\lambda$4363 emission line. Particularly,
the modified relation (Eq.~\ref{eq2}) should be used for the rare
galaxies with extremely high O$_{32}$ ratios above $\sim$40, whereas
the simpler relation derived by \citet{Iz19b} can be applied to the vast
majority of SFGs with lower O$_{32}$.

\section{Conclusions}\label{sec:conclusions}

We present Sloan Digital Sky
Survey (SDSS) and 3.5m Apache Point Observatory (APO) Telescope/Kitt Peak Ohio
State Multi-Object Spectrograph (KOSMOS) spectrophotometric observations
of the compact star-forming galaxy (SFG)
J1046$+$4047, selected from the SDSS Data Release 16 (DR16). This local SFG
possesses a very low metallicity and an extremely high O$_{32}$ ratio. Our main results are as follows.

1. The emission-line spectrum of J1046$+$4047, with a stellar mass of 
1.8$\times$10$^6$ M$_\odot$ and an absolute SDSS $g$-band magnitude 
of --16.51 mag, originates from a hot [$T_{\rm e}$(O~{\sc iii})~=~25300~K]
ionised gas.
The detection of the [Fe~{\sc v}] $\lambda$4227 and He~{\sc ii} $\lambda$4686
emission lines indicates the presence of hard ionising radiation.
The oxygen abundance in J1046$+$4047 is 12+log(O/H) = 7.082$\pm$0.016, 
one of the lowest for nearby SFGs. 

2. The rest-frame equivalent width EW(H$\beta$) of the H$\beta$ 
emission line of 411\AA\ in the J1046$+$4047 spectrum is one of the highest
measured so far in SFGs. Both the very high EW(H$\beta$) and very 
low $M_\star$/$L_g$
indicate that the J1046$+$4047 emission is strongly dominated by a very young 
stellar population with an age $\la$ 2 Myr. These properties are similar to those 
of another 
very metal-deficient SFG known, J2229$+$2725
\citep{Iz21}. 

3. The O$_{32}$ = $I$([O~{\sc iii}]$\lambda$5007)/$I$([O~{\sc ii}]$\lambda$3727)
flux ratio of $\sim$ 57 in J1046$+$4047 is extremely high and is similar
to the O$_{32}$ = 53 of J2229$+$2725 with nearly
the same oxygen abundance. These properties together with
the very high rest-frame equivalent width EW(H$\beta$) of the H$\beta$
emission line imply that both galaxies 
are at a very young stage of their formation. The unique properties of
J2229$+$2725 and J1046$+$4047 likely make them the best local counterparts of
young high-redshift dwarf SFGs with stellar
masses and metallicities similar to those of the galaxies,  
recently discovered at $z$ $\sim$ 6 -- 8 by \citet{A23}.

4. The extremely high O$_{32}$ ratio in J1046$+$4047 allows us to verify and
slightly improve the strong-line method by \citet{Iz21}, based on the intensities of 
strong nebular [O~{\sc ii}]$\lambda$3727 and 
[O~{\sc iii}]$\lambda$4959+$\lambda$5007 emission lines. We can 
use it for the
determination of the oxygen abundance in extremely low-metallicity SFGs, in the range 
12~+~log(O/H)~$\la$~7.65 and O$_{32}$~$\la$~60. 

5. We find that the H$\alpha$ emission line in J1046$+$4047 is enhanced by
some non-recombination processes such as, for example, collisional excitation of
H$\alpha$ in the circumstellar envelopes of luminous massive stars or in dense
supernova remnants. Therefore, this line can not be used for interstellar
extinction determination in J1046$+$4047.

\section*{Acknowledgements}

Y.I.I. and N.G.G. acknowledge support from the National Academy of Sciences of
Ukraine (Project No. 0123U102248) and from the Simons Foundation.
This study is based
on observations with the the 3.5m Apache Point Observatory (APO). The Apache 
Point Observatory 3.5-meter telescope is owned and operated by the 
Astrophysical Research Consortium.
{\sc iraf} is distributed by the 
National Optical Astronomy Observatories, which are operated by the Association
of Universities for Research in Astronomy, Inc., under cooperative agreement 
with the National Science Foundation.
Funding for the Sloan Digital Sky Survey IV has been provided by
the Alfred P. Sloan Foundation, the U.S. Department of Energy Office of
Science, and the Participating Institutions. SDSS-IV acknowledges
support and resources from the Center for High-Performance Computing at
the University of Utah. The SDSS web site is www.sdss.org.
SDSS-IV is managed by the Astrophysical Research Consortium for the 
Participating Institutions of the SDSS Collaboration. 
Based on observations 
made with the NASA Galaxy Evolution Explorer. GALEX is operated for NASA by 
the California Institute of Technology under NASA contract NAS5-98034. 
This research has made use of the NASA/IPAC Extragalactic Database (NED), which 
is operated by the Jet Propulsion Laboratory, California Institute of 
Technology, under contract with the National Aeronautics and Space 
Administration.

\section*{Data availability}

The data underlying this article will be shared on reasonable request to the 
corresponding author.

\bsp

\label{lastpage}


\begin{thebibliography}{}




\bibitem[Ahumada et al.(2020)]{Ah20} Ahumada R. et al., 2020, \apjs, 249, 1







\bibitem[Annibali et al.(2019)]{An19} Annibali F. et al., 2019, \mnras, 
482, 3892


\bibitem[Atek et al.(2023)]{A23} Atek H. et al., 2023,
preprint arXiv:2308.08540

\bibitem[Baldwin et al.(1981)Baldwin, Phillips \& Terlevich]{BPT81} 
Baldwin J. A., Phillips M. M., Terlevich R., 1981, \pasp, 93, 5





\bibitem[Berg et al.(2021)]{B21} Berg D. A., Chisholm J., Erb D. K.,
Skillman E. D., Pogge R. W., Olivier G. M., 2021, \apj, 922, 170







\bibitem[Bouwens et al.(2015)]{B15} Bouwens R. J., Illingworth G. D., 
Oesch P. A., Caruana J., Holwerda B., Smit R.,  Wilkins S., 
2015, \apj, 811, 140



















\bibitem[Filippenko(1982)]{F82} Filippenko A. V., 1982, \pasp, 
94, 715





\bibitem[Fricke et al.(2001)]{F01} Fricke K. J., Izotov Y. I.,
Papaderos P., Guseva N. G., Thuan T. X., 2001, \aj, 121, 169


\bibitem[Garnett(1992)]{G92} Garnett D. R., 1992, \aj, 103, 1330



\bibitem[Girardi et al.(2000)]{G00} Girardi L., Bressan A., Bertelli G., 
Chiosi C., 2000, \aaps, 141, 371




\bibitem[Guseva et al.(2012)]{G12} Guseva N. G., Izotov Y. I., Fricke K. J., 
Henkel C., 2012, \aap, 541, A115




\bibitem[Heintz et al.(2022)]{He22} Heintz K. E. et al., 2022, preprint
arXiv:2212.02890

\bibitem[Hirschauer et al.(2016)]{H16} Hirschauer A. S. et al., 2016, \apj, 822,
108







\bibitem[Izotov \& Thuan(2009)]{IT09} Izotov Y. I., Thuan T. X., 
2009, \apj, 707, 1560


\bibitem[Izotov et al.(1994)Izotov, Thuan \& Lipovetsky]{ITL94} Izotov Y. I.,
Thuan T. X., Lipovetsky V. A., 1994, \apj, 435, 647

\bibitem[Izotov et al.(2001)Izotov, Chaffee \& Schaerer]{I01} Izotov Y. I.,
Chaffee F. H., Schaerer D., 2001, \aap, 378, L45



\bibitem[Izotov et al.(2006)]{I06} Izotov Y. I., Stasi\'nska G., Meynet G.,
Guseva N. G., Thuan T. X., 2006, \aap, 448, 955




\bibitem[Izotov et al.(2009)]{I09} Izotov Y. I., Guseva N. G., Fricke K. J., 
Papaderos P., 2009, \aap, 503, 61


\bibitem[Izotov et al.(2012)Izotov, Thuan \& Guseva]{I12} Izotov Y. I., 
Thuan T. X., Guseva N. G., 2012, \aap, 546, 122













\bibitem[Izotov et al.(2018a)]{I18a} Izotov Y. I., Thuan T. X., Guseva N. G., 
Liss S. E., 2018a, \mnras, 473, 1956 

\bibitem[Izotov et al.(2018b)]{I18b} Izotov Y. I., Schaerer D., Worseck G., 
Guseva N. G., Thuan, T. X., Verhamme A., Orlitov\'a I., Fricke K. J.,
2018b, \mnras, 474, 4514 

\bibitem[Izotov et al.(2018c)]{I18c} Izotov Y. I., Worseck G., Schaerer D.,
Guseva N. G., Thuan, T. X., Fricke K. J., Verhamme A., Orlitov\'a I., 
2018c, \mnras, 478, 4851 

\bibitem[Izotov et al.(2019a)Izotov, Thuan \& Guseva]{Iz19a} Izotov Y. I.,
Thuan T. X., Guseva N. G., 2019a, \mnras, 483, 5491 

\bibitem[Izotov et al.(2019b)]{Iz19b} Izotov Y. I., Guseva N. G., Fricke K. J.,
Henkel C., 2019b, \aap, 623, 40


\bibitem[Izotov et al.(2021a)Izotov, Thuan \& Guseva]{Iz21} Izotov Y. I.,
Thuan T. X., Guseva N. G., 2021a, \mnras, 504, 3996 

\bibitem[Izotov et al.(2021b)Izotov, Thuan \& Guseva]{Iz21b} Izotov Y. I., 
Thuan T. X., Guseva N. G., 2021b, \mnras, 508, 2556


\bibitem[Jaskot \& Oey(2013)]{JO13} Jaskot A. E., Oey M. S.,
2013, \apj, 766, 91





\bibitem[Kennicutt(1998)]{K98} Kennicutt R. C., Jr.,
1998, \araa, 36, 189

\bibitem[Kojima et al.(2020)]{Ko20} Kojima T. et al., 2020, \apj, 898, 142








\bibitem[Leitherer et al.(1999)]{L99} Leitherer C. et al., 1999, \apjs, 123, 3



\bibitem[Lejeune et al.(1997)Lejeune, Buser \& Cuisiner]{L97} 
Lejeune T., Buser R., Cuisinier F., 1997, \aaps, 125, 229









\bibitem[Mitra et al.(2013) Mitra, Ferrara \& Choudhury]{M13} 
Mitra S., Ferrara A., Choudhury T. R., 2013, \mnras, 428, L1





\bibitem[Ouchi et al.(2009)]{O09} Ouchi M. et al., 2009, \apj, 706, 1136





\bibitem[Planck Collaboration XVI(2014)]{P14} Planck Collaboration XVI,
2014, \aap, 571, 16



\bibitem[Press et al.(2007)]{Pr07} Press W. H., Teukolsky S. A.,
Vetterling W. T., Flannery B. P., 2007, Numerical Recipes 3rd Edition: The Art
of Scientific Computing, New York: Cambridge University Press, 1235 pp.













\bibitem[Salpeter(1955)]{S55} Salpeter E. E., 1955, \apj, 121, 161

\bibitem[Saxena et al.(2023)]{Sa23} Saxena A. et al., 2023, preprint
arXiv:2306.04536














\bibitem[Storey \& Hummer(1995)]{SH95} Storey P. J., Hummer D. G., 
1995, \mnras, 272, 41


\bibitem[Thuan \& Izotov(2005)]{TI05} Thuan T. X., Izotov Y. I., 2005, \apjs,
161, 240


\bibitem[Thuan et al.(2022)Thuan, Guseva \& Izotov]{T22} Thuan T. X. Guseva N. G., Izotov Y. I., 2022, \mnras, 516, L81










\bibitem[Wise \& Chen(2009)]{WC09} Wise J. H., Cen R.,
2009, \apj, 693, 984


\bibitem[Yajima et al.(2011) Yajima, Choi \& Nagamine]{Y11} 
Yajima H., Choi J.-H., Nagamine K., 2011, \mnras, 412, 411







\end{thebibliography}
\end{document}